\documentclass{aa}
\usepackage{graphicx}
\long\def\symbolfootnote[#1]#2{\begingroup%
 \def\thefootnote{\fnsymbol{footnote}}\footnote[#1]{#2}\endgroup} 
\begin{document}
\title{\bf A 1.2 mm MAMBO survey of Post-AGB stars}
\author{      C. S. Buemi\inst{1}
          \and G. Umana \inst{1}
         \and C. Trigilio \inst{1}
	 \and P. Leto \inst{2}
                      }
\institute{INAF,  Osservatorio Astrofisico di Catania
              via S. Sofia, 78 
	      95123, Catania Italy
   \and
   INAF, Istituto di Radioastronomia
              Sezione di Noto, C.P. 161 96017, Noto (SR) Italy\\
           }
\offprints{C. Buemi, cbuemi@oact.inaf.it}
\date{Received \hbox to1in{\hrulefill}}

 \abstract{}
{We performed a millimetric survey of a sample of 24 post-AGB stars aimed to search for emission from
circumstellar matter, in order to investigate the physical properties of the outer parts of the
envelopes.}
{The observations were conducted using the 37-channel Max-Planck Millimeter Bolometer array
at the 30-meter IRAM telescope. The continuum emission toward the detected sources was used to quantify the mass
of the emitting dust. We combined our observations with data available in literature to construct the 
spectral energy distribution (SED) of the sources. When the observational data cover a spectral range wide enough,
some properties of circumstellar envelopes were derived by comparison with spectra computed using a radiative transfer code.}
{Of the 24 objects in our sample, we detected millimetric continuum emission toward 11 sources. Two other sources
were detected at flux level close to 3$\sigma$. The derived circumstellar dust masses range between 0.4 and 
$24 \times 10^{-4}$ M$_{\odot}$, but these results are affected from the uncertain about the source distances.
The parameters derived from the SED fits are consistent with the values characteristic for these kind of object. As confirmed from the flux density extrapolated in the first light channels of the Atacama Large Millimetric Array, such sources would be good targets for future high resolution mapping with the ALMA facility.}{}
\authorrunning{Buemi  et al.}
\maketitle
%
\section{Introduction} 
Stars on the red giant branch undergo a process of copious mass-loss
leading to the formation of a circumstellar envelope (CSE).
These envelopes, quilted up by the ejected material, contain atomic 
and molecular gas and are characteristic of the post Asymptotic Giant Branch (AGB) evolutionary phase,
which results in the formation of a Planetary Nebula (PNe). 
Because of the huge amount of processed material returned to the ISM,  this evolutionary phase is very important 
for the chemical evolution of the Galaxy. 
Yet, the short transition phase between the end of the AGB and the formation of a new PNe is still 
poorly understood. In particular, it is quite challenging to understand how the almost symmetric CSE observed 
around AGB stars transform themselves in the highly structured morphologies observed in high dynamical 
range optical images of PNs.

The importance of studying objects  in the way to PNe  resides in the fact that in their CSE
the unknown physical mechanisms, that shape the PNe,  are already  at work,
as HST images of multi-polar outflows in post-AGB stars appear to indicate (Sahai \cite{sahai01}).

Despite of numerous efforts to identify the shaping agent/s in PNs
(Sahai \& Trauger \cite{sahai98}; Garcia-Segura et al. \cite{garciaseg}), there are no observative evidences that support one mechanism in favors to the others  (Balick and Franck \cite{balick}). 

Dust is ubiquitous detected in post-AGB and PNe and, quite often, it is  in the form of 
disks and tori. Some mo\-dels evocate these structures,  whose origin is however under debate,
as important ingredient for the mechanism which produces collimated outflows
observed in the CSE (Frank et al. 1997; Huggins et al. \cite{Huggi}).
In this contest, very important informations can be provided by stu\-dying the physical properties 
of CSE and in particular, the pre\-sen\-ce and the spatial distribution of the circumstellar dust, 
as this may help in finding clues to understand the nature of the shaping agent.

Detailed studies of the spectral energy distribution (SED) of a small sample of
optically bright post-AGB candidate stars have shown that
these objects can be divided into two groups, depending on the shape of
the IR excess (Trams et al. \cite{trams}; van der Veen et al. \cite{vanderveen}):
sources with a broad IR excess extending from the near infrared
until the far-IR have both hot and cool dust in their circumstellar
shells and sources with only a far-IR  excess show only the presence
of cool dust. The double peaks in the SEDs appear to be characteristic for objects 
in transitions, but the presence of only cool dust seems to point out
objects more evolved towards PNs.
\begin{table*}
\caption[]{Properties of stars  in our sample. {\it IRAS} fluxes come directly from Point
Source Catalog (PSC).}
\label{sample}
\begin{center}
\begin{tabular}{ l c c rrrr}
\multicolumn{7}{c}{Sources with Optical Counterpart} \\
  \hline
   \hline
           &             &            & & & &\\
IRAS Name  & RA(J2000)   & DEC(J2000) & F$_{12}$ & F$_{25}$ & F$_{60}$ & F$_{100}$ \\
           &             &            &   [Jy]   &    [Jy]   &   [Jy]    &  [Jy]       \\  
\hline
00210+6221 & 00 23 51.2  & +62:38:16  & 48.5 & 51.9  & 12.5 & $<$23.2\\
01174+6110 & 01:20:44.9  & +61:26:18  &  4.1 & 16.9  & 33.9 &  4.1\\ 
04296+3429 & 04:32:56.6  & +34:36:11  & 12.7 & 45.9  & 15.4 & $<$9.2\\ 
05089+0459 & 05:11:36.1  & +05:03:26  &  7.4 & 21.9  & 11.9 &  3.8\\ 
06530--0213 & 06:55:32.1  & $-$02:17:30  &  6.1 & 27.4  & 15.1 &  4.1\\ 
07134+1005 & 07:16:10.2  & +09:59:48  & 24.5 & 116.7 & 50.1 & 18.7\\ 
07331+0021 & 07:35:41.1  & +00:14:58  & 15.3 & 68.1  & 18.5 &  3.7\\ 
17436+5003 & 17:44:55.4  & +50:02:39  &  6.1 & 184.0 &152.0 & 48.7\\ 
19114+0002 & 19:13:58.6  & +00:07:31  & 31.3 & 648.3 &515.9 &168.1\\
20000+3239 & 20:01:59.4  & +32:47:32  & 15.0 &  71.0 & 30.0 & $<$43.1 \\
20028+3910 & 20:04:35.0  & +39:18:38  & 41.8 & 210.8 &143.1 & 46.5\\ 
22223+4327 & 22:24:30.6  & +43:43:03  &  2.1 &  37.1 & 22.4 &  9.5\\ 
22272+5435 & 22:29:10.3  & +54:51:06  & 73.9 & 302.4 & 96.6 & 41.0\\ 
22574+6609 & 22:59:18.4  & +66:25:48  &  9.0 &  29.5 & 10.7 &  2.5\\
23304+6147 & 23:32:45.0  & +62:03:49  & 11.4 &  59.1 & 26.6 &  7.2\\
\hline
           &             &            & & & &\\
\multicolumn{7}{c}{Sources without Optical Counterpart} \\
  \hline
   \hline
07430+1115 & 07:45:49.8  & +11:08:25  &  7.7 & 22.9 & 10.7 & 2.5\\
18454+0001 & 18:48:01.5  & +00:04:47  & 10.8 & 14.5 & 13.6 &$<$384\\
18514+0019 & 18:53:57.9  & +00:23:24  &  4.9 & 23.4 & 17.3 &$<$152\\
18576+0341 & 19:00:11.2  & +03:45:46  & 58.5 &425.0 &274.7 &$<$1660\\
19024+0044 & 19:05:01.5  & +00:48:48  &  2.9 & 48.8 & 42.5 & 15.7\\
19075+0432 & 19:10:00.0  & +04:37:06  &  5.2 & 28.1 & 31.8 & 14.4\\
19454+2920 & 19:47:24.3  & +29:28:12  & 17.3 & 89.6 & 54.4 & 14.7\\
20144+4656 & 20:15:58.3  & +47:05:39  &  1.2 & 17.0 & 20.0 & $<$85\\
21537+6435 & 21:55:04.6  & +64:49:54  &  6.9 & 26.1 & 13.3 & $<$6.1\\
\hline
\end{tabular}
\end{center}
\end{table*}
Thermal emission from dusty envelopes may extend up to millimiter
wavelengths, depending on the temperature of the dust. 
In particular, millimetric observations are essential to better determine the SEDs of the sources and thus 
to put more stringent constraints  to the model  of their circumstellar envelopes. Moreover, millimetric observations
will allow to assest the presence of multiple dusty shells to be related to different mass-loss episodes undergone by the stars.

In this paper we present the results of a 1.2 mm survey aimed to detect the millimetric emission from a sample of post-AGB stars
for a robust modeling of their circumstellar envelopes. This appears as a first, necessary step in preparation for the use of the
new interferometers (i.e. ALMA and CARMA) that, in the very near future, will be able to resolve, in unprecedented spatial resolution, the circumstellar geometry
for a full understanding of the shaping mechanisms.

\section{Observations}
\subsection{Sample selection}

In the attempt to determine a possible millimetric  emission
due to thermal dust emission from the circumstellar envelope,
we observed a sample of Post-AGB stars with the 30m IRAM telescope.

Since the dusty envelopes show a clear signature in the far-IR
spectrum of the stars, the IRAS colour-colour
diagram has been successfully used by several
authors in systematic searches for post-AGB objects by looking
for sources in-between the locus of the planetary nebulae and
late-type AGB stars. Garcia-Lario et al. (\cite{garcia}) compiled a list of stellar objects, 
characterized by strong F-IR excess, which occupy the same region
of the  IRAS color dia\-gram as AGB stars and PNs. 
The sample contains 126 Post-AGB. Among those, a large fraction ($70\%$) represents 
post-AGB in a very early stage, still heavily obscured in the optical.

Another method of looking  for transition objects is concentrated
on optically bright objects with an IR excess due to circumstellar dust
(Pottash and Parthasarathy \cite{Pottasch}; Trams et al. \cite{trams}; Oudmaijer et al. \cite{oud92}).
These resulted in the detection of objects scattered in the IRAS
colour-colour diagram, as they show different amounts of
IR excesses and different IR-colours.

Oudmaijer et al (\cite{oud92}; \cite{oud96}), looked for transition objects with an optical
counterpart by performing a cross-correlation of the SAO optical catalogue
with the IRAS point-source catalogue, selecting supergiants with
spectral type between B and G and with an IR excess due to circumstellar dust.

To select a sample of possible targets we used the most complete compilations of 
objects with no optical counterpart, from Garcia-Lario et al. (\cite{garcia})  
and of objects with optical counterpart, 
from Oudmaijer et al (\cite{oud92}; \cite{oud96}), selecting the stars identified as
post-AGB, which are associated with highly evolved post-AGB stars with low-mass progenitors. We further include in 
the sub-sample of stars with optical counterpart 11 stars, originally not classified by Oudmjier et al.
(\cite{oud92}) as Post-AGB,  for which  there are compelling
observational evidences that are in the post-AGB evolutionary stage
(van der Veen et al. \cite{vanderveen}; van Wincker \cite{vanwin97}).
\begin{table}[]
\caption[]{Measured millimeter fluxes.}
\label{flux}
\begin{center}
\begin{tabular}{ l c r  r  c }
\multicolumn{5}{c}{Sources with Optical Counterpart} \\
  \hline
  \hline
           &             &           &		&	\\
IRAS Name  &    Date     &  Time     &	Flux~  &  Weather$^{\dag}$\\
           &             &  [s] ~~& [mJy]	  &	\\  \hline
00210+6221 & 21 Jan 02  & 1200 &	$\dots\pm$2.4	& B		\\
01174+6110 & 21 Jan 02  & 1200 &	22.9$\pm$1.7	& B		\\
04296+3429 & 14 Feb 02  & 1200 &	4.4$\pm$1.3	& A		\\
05089+0459 & 05 Feb 02  & 1440 &	$\dots\pm$1.0	& A		\\
06530$-$0213 & 14 Feb 02  & 1200 &	4.8$\pm$1.5	& A	\\
07134+1005 & 07 Feb 02  & 1200 &	14.0$\pm$1.5	& A		\\
07331+0021 & 07 Feb 02  & 1200 &	$\dots\pm$1.9	& B		\\
17436+5003 & 05 Feb 02  & 1080 &	15.2$\pm$1.1	& A		\\
19114+0002 & 04 Feb 02  & 480 &	68.2$\pm$3.4	& B		\\
20000+3239 & 21 Jan 02  & 1200 &	11.4$\pm$1.7	& B	\\
20028+3910 & 21 Jan 02  & 1200 &	11.9$\pm$1.5	& A	\\
22223+4327 & 04 Feb 02  & 1200 &	$\dots\pm$1.5	& A		\\
22272+5435 & 04 Feb 02  & 1200 &	35.3$\pm$1.7	& A		\\
20144+4656 & 21 Jan 02  & 1200 &	$\dots\pm$1.8	& B		\\
23304+6147 & 21 Jan 02  & 1200 &	$\dots\pm$1.8	& B		\\
\hline
 & & & &\\
\multicolumn{5}{c}{Sources without Optical Counterpart} \\
  \hline
  \hline
07430+1115 & 07 Feb 02  & 1200 &	$\dots\pm$1.3	& A \\
18454+0001 & 22 Jan 02  & 1200 &	$\dots\pm$1.4	& A	\\
18514+0019 & 21 Jan 02  & 1200  &  8.2$\pm$1.3  & A \\
18576+0341 & 03 Feb 02  & 840  & $\ge$45.4$\pm$2.2   &      B       \\
19024+0044 & 22 Jan 02  & 1200 &	$\dots\pm$1.3	& A		\\
19075+0432 & 04 Feb 02  & 1200  &	6.2$\pm$1.2   & A		\\
19454+2920 & 03 Feb 02  & 1200  &	$\dots\pm$4.3   &	C	\\
20144+4656 & 05 Feb 02  & 1200 &	$\dots\pm$2.7	& C		\\
21537+6435 & 04 Feb 02  & 1200  &	6.3$\pm$1.6   &	A	\\
\hline
\multicolumn{5}{l}{\footnotesize{$^{\dag}$ Wheatear condition: A) good, B) poor, C) bad.}} \\
\end{tabular}
\end{center}
\end{table}

A small sample of post-AGB stars has been observed in the millimetric band
with the JCMT (van der Veen et al. \cite{vanderveen}) and the detection rate appeared to be
well correlated with the F$_{60}$ IRAS flux, in the sense that all the stars
with F$_{60}$ $ \geq$ 10 ~Jy have been detected. If  the extra infrared excess has the shape of a cold black body, 
as expected from a more distant dust shell,
a F$_{60}$ $\sim$ 10 Jy would imply a 1.3 mm flux higher than 36 mJy and thus
easily detectable with the new Bolometer.
In order to maximize the probability to have a detectable flux  at 1.2 mm
we thus selected, from our original sample, only those sources with
${\mathrm F}_{60} \geq 10$ Jy. This reduces our sample to 34 targets.\par
In this paper we report on the results relative to  a subsamples of 24 objects,
which are given in Table~\ref{sample}.

\subsection{The 30m IRAM observations and results}

The 37-channel Max-Planck Millimeter Bolometer (``MAMBO''; Kreysa et al. \cite{kreysa}) array
at the 30-meter IRAM telescope on Pico Veleta (Spain) was used to perform the survey.
The observations were made between 21 January and 7 February 2002, using the standard
 ON-OFF technique, chopping the secondary mirror of the telescope by about 50$\arcsec$ in azimuth,
at a rate of 1 Hz. The FWHP of our beam was 10$\arcsec$.5 at 1.2 mm.

For each source the observations were typically obtained in blocks of 4 or 5 scans lasting 4 minutes each.
Frequent skydip observations have been used to determine atmospheric extinction as function of elevation and time.
The data were analyzed using the MOPSI software (Zylka \cite{zylka}).
The flux calibration was performed by observing either Mars and Uranus to determine the
flux conversion factor. For each channel the sky
noise was subtracted by computing the weight mean of the signals from the surrounding
six channels. 

\begin{figure*}
\resizebox{17cm}{!}{\includegraphics{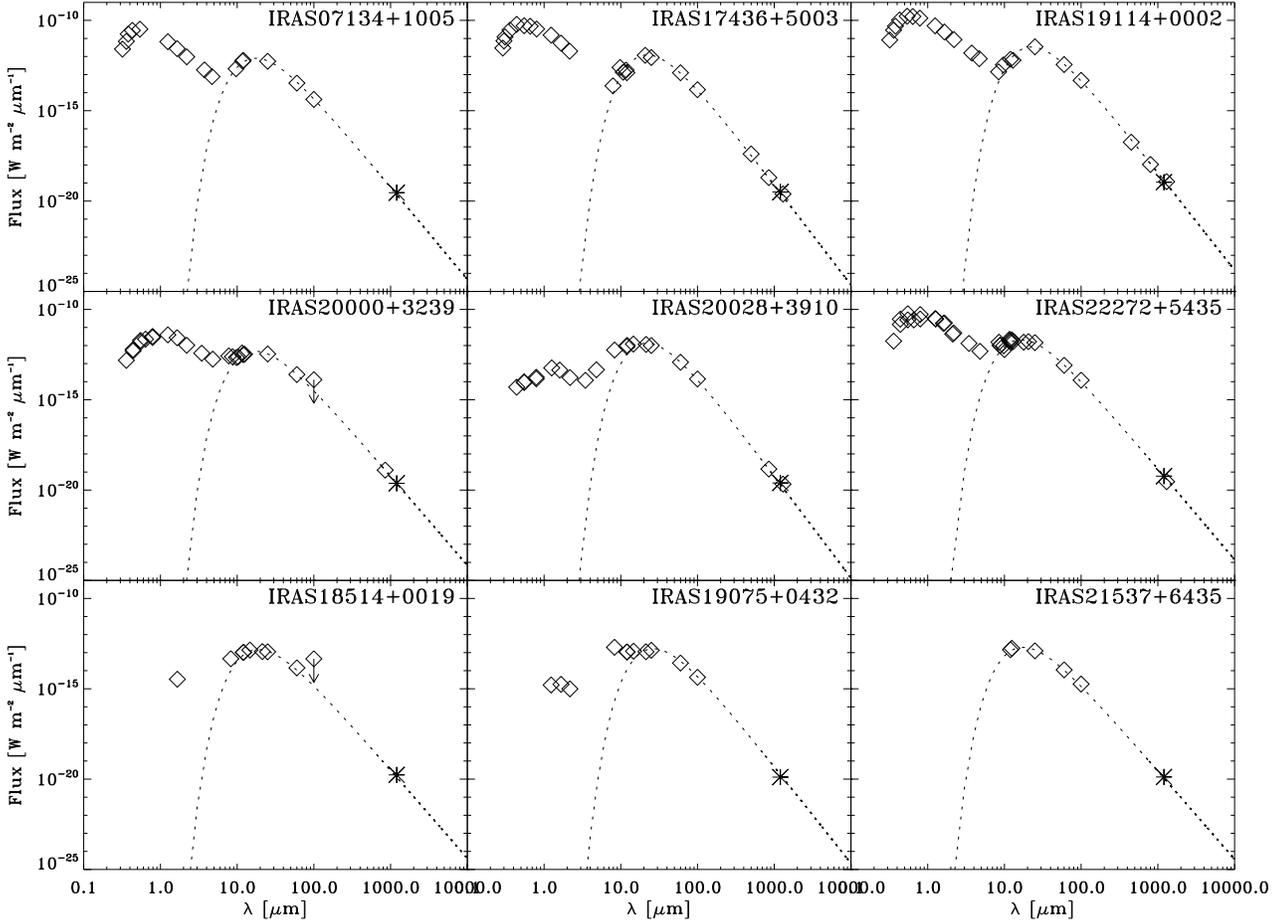}}
\caption{Spectral energy distributions of the emission towards the detected sources.
The asterisks indicate measures from this work.
The dashed lines show the modified blackbody fits to the long-wavelength data, as discussed in the text (par. 3.1). }
\label{bbody}
\end{figure*}

%
%
In table \ref{flux} the resulting 1.2 mm flux densities measured for the detected objects are reported.
Out of the 24 observed sources,
we carried out 11 detections over the threshold of $4\sigma$ and two uncertain detections at $3\sigma$.
Four of the detected and one of the undetected sources overlap the sample observed from Walmsley et al. (\cite {walmsley}),
by using the 30m IRAM telescope at the wavelength of 1.3~mm, but with a different bolometer. 
The measured flux densities are in  good accordance
with the exception of IRAS~22272+5435, which shows a flux level significatively higher.

The sample includes IRAS~01174+6110 and IRAS~18576+0341, which were firstly classified as post-AGB stars on the basis of their IRAS colours, but probably having different nature. IRAS~01174+6110, in fact, is likely an HII region
(Kelly et al. \cite{kelly}), while IRAS~18576+0341 has been recently recognized as new LBV (Pasquali $\&$ Comeron \cite{pasquali}; Clark et al. \cite{clark}; Umana et al. \cite{Umana}). For these two objects we just report the measured flux density. \par
In particular, IRAS~18576+0341 will be object of a future more detailed analysis assembled with radio high resolution results. It is to note that Umana et al. (\cite{Umana}) pointed out that the source's right ascension is about 4$\arcsec$ shifted with respect to values previously reported in li\-te\-ra\-tu\-re (Garcia-Lario \cite{garcia}). Since the first contour level in the 22 GHz map defines a source size close to 10$\arcsec$, it is reasonable to consider the measured 1.2 mm flux as a lower limit.

 Some of the detected object in our sample have been detected in $^{13}$CO and/or $^{12}$CO transition J=2-1
at $\lambda$= 1.3 mm, that is within the band of our observations (Hrivnak et al. \cite{hriv05}, Bujarrabal et al. \cite{bujar92}, Bujarrabal et al. \cite{bujar01}).
We verified that the contribution of such emission, spreaded in our bandwidth of about
80 GHz, is lower than the errors associated to the measures for all the detected sources but IRAS~19114+0002 and
IRAS~22272+5435. In particular, on the basis of the observations by  Bujarrabal et al. (\cite{bujar92}), we derived
for IRAS~19114+0002 an emission, in the MAMBO band, of 15 mJy from $^{12}$CO and of 1.6 mJy from  $^{13}$CO;
for IRAS~22272+5435 the contribution of $^{12}$CO emission is about 7 mJy (Hrivnak et al. \cite{hriv05}). In the following calculations, we thus subtracted these contributions from the observed fluxes.

\subsection{Spectral energy distributions}

In order to build up the spectral energy distribution (SED) of the detected sources, 
updated with our millimetric data,
and to investigate its corrected low frequencies shape, we combined our 
MAMBO observations
with infrared (2MASS + IRAS + MSX) and optical data available in literature. The data are 
corrected for the interstellar extinction, with the exception of  
IRAS~19075+0432 due to the lack of informations about the extinction toward the source. Moreover we have excluded from this analysis IRAS~01174+6110
because of the uncertainty on its nature.

For those objects for which the observations cover a spectral range wide enough,
the resulting SEDs (fig.~\ref{bbody}) show 
the typical double peaked shape, with the two peaks corresponding respectively
to the optical emission from the  photosphere of the central star and to the thermal 
emission from the circumstellar dust.

\begin{table*}
\caption[]{Derived values for the emissivity index, dust mass, absorption coefficient and envelope dust mass.}
\label{masstab}
\begin{center}
\begin{tabular}{ l c c c  r  c }
\multicolumn{6}{c}{Source with Optical Counterpart} \\
  \hline
  \hline
          &	 &             &           &		&    		\\
IRAS Name &  T$_d$   &  p   & $\chi_{1.3}$  & M$_{d}$ ~~ & d ~~\\
          & [K]      &      & $[cm^2 g^{-1}]$ &	 [M$_{\odot}$]~~ & [kpc] \\  
	   \hline
07134+1005 & 135 & 1.11  & 1.75 & 8.5 (--4)	&  2.4    \\
17436+5003 & 100 & 1.55  & 0.88 & 6.2 (--4)	&  1.2    \\
19114+0002 & 100 & 1.41 & 1.09 & 4.3 (--2)	&  6.0      \\
20000+3239 & 140  & 0.94 -- 1.55   & 0.88 -- 2.1 & 0.9 -- 2.3 (--4)	&  1.0$^{\dag}$\\
20028+3910 & 100 & 1.65  & 0.75 & 2.4 (--3)	&  2.9    \\
22272+5435 & 145 & 1.03  & 1.98  & 6.2 (--4)	&  1.6    \\
\hline
          &	 &             &           &		&    		\\
\multicolumn{6}{c}{Source without Optical Counterpart} \\
\hline
  \hline
18514+0019 & 130 & 0.89 -- 2.18   & 0.32 -- 2.5 & 0.6 -- 4.8 (--4) &  1.0$^{\dag}$    \\
19075+0432 & 85  & 1.37  & 1.2  &  1.5 (--4)   &  1.0$^{\dag}$ \\
21537+6435 &140  & 0.87 -- 0.97   & 2.18 -- 2.55 & 4.3 -- 5.1 (--5) &  1.0$^{\dag}$       \\
\hline
\multicolumn{6}{l}{\footnotesize{$^{\dag}$ Assumed distance.}} \\
\end{tabular}
\end{center}
\end{table*}

\section{Analysis}
\subsection{Dust Masses}

From a theoretical point of view, the mass of the circumstellar dust surrounding the central object can be derived
from the observed continuum millimetric flux. Assuming that the flux density measured at this frequencies is to be
ascribed to the thermal emission from optically thin and isothermal dust, the dust mass ($M_\mathrm{d}$) 
and the flux density are directly proportional (Hildebrand \cite{hildeb}):

\begin{displaymath}
M_\mathrm{d}= \frac{F(\nu)d^2}{B_{\nu}(T_\mathrm{d}) \chi_{\nu}}
\end{displaymath}

\noindent
where $B_\nu(T_\mathrm{d})$ is the Planck function for dust temperature $T_\mathrm{d}$, $d$ is the distance to the source and $\chi_{\nu}$ is
the dust opacity at the observing frequency.
We can thus estimate the dust mass by using the Rayleigh-Jeans approximation:

\begin{equation}
M_\mathrm{d}= \frac{F(\nu) \lambda^2 d^2}{2kT_\mathrm{d} \chi_{\nu}}.
\label{eq1}
\end{equation}
The value of $\chi_{\nu}$ is the major uncertainty that affects the conversion of the millimetric flux
density in dust mass. Following Hildebrand's approach, we can extrapolate the dust opacity $\chi_{\nu}$ at 1.2 mm
from its value at 250 $\mu$m, i.e. $\chi_{250\mu \mathrm{m}} = 10$ cm$^2$ g$^{-1}$, assuming the power-law dependence 
$\chi_{\nu}\propto\nu^p$, where $p$ is the emissivity index and strongly depends on the mineralogical composition
of the grain and on their physical shape.

Under the hypothesis of optically thin emission, the emissivity index $p$ may be derived from the spectral 
index in the millimetric and submillimetric spectral range, where 
the dust emits as a blackbody modified by the frequency dependent dust opacity, that is $F_{\nu}\propto \chi_{\nu}B_{\nu}(T_\mathrm{d})$.

We thus proceeded iteratively by fitting a modified blackbody to the infrared and millimetric data to estimate
the dust temperature, while the emissivity index has been derived from a linear fit in the $\log{\nu}$ - $\log\frac{F_{\nu}}{B(T_\mathrm{d})}$ to the integrated flux density from 100 $\mu$m to 1.2 mm. In the case of IRAS~19075+0432, it is to note that the MSX data at 8.28 $\mu$m suggest the presence of both warm and cold circumstellar dust.
\begin{figure*}
\resizebox{17cm}{!}{\includegraphics{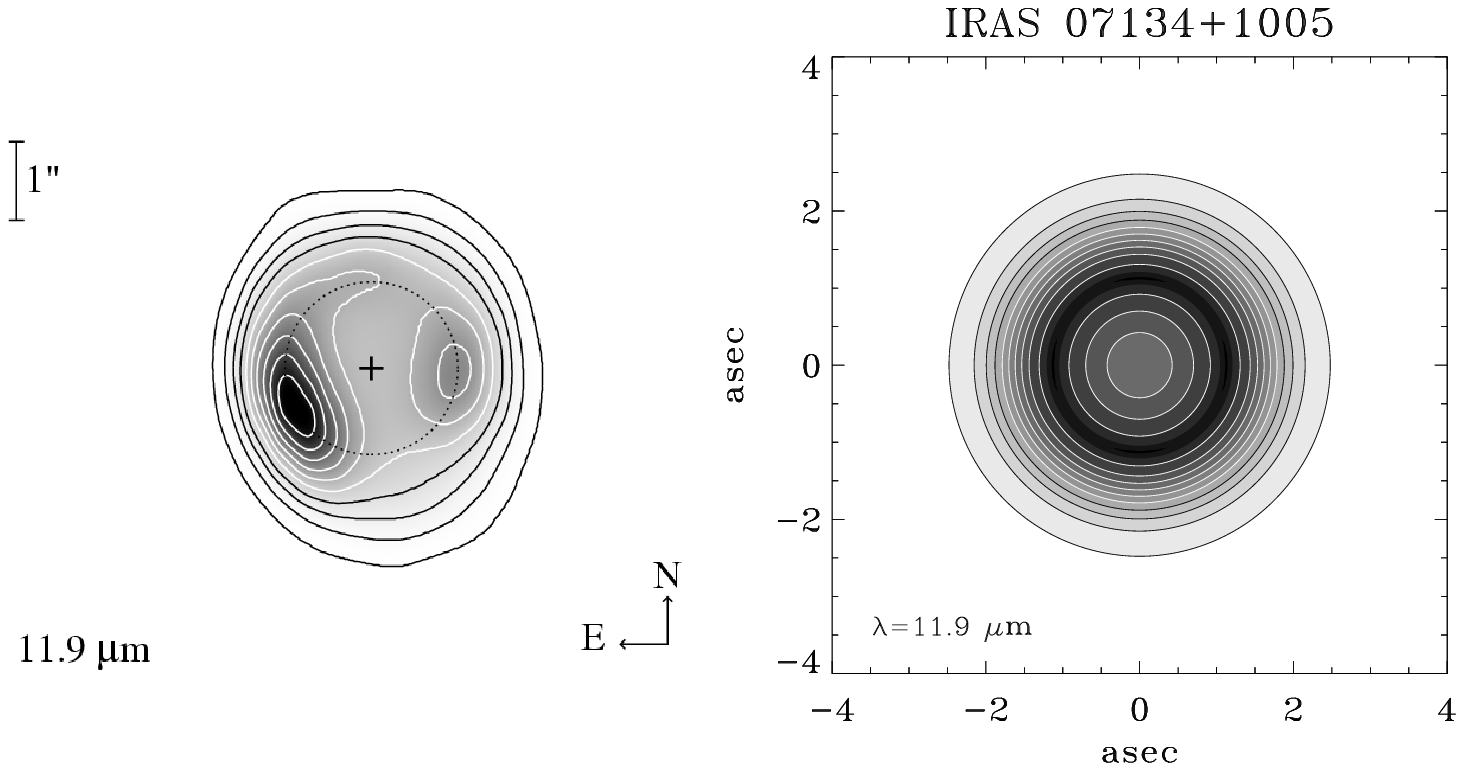}}
\caption{{\it Left:} Image of IRAS~07134+1005 observed at 11.9 $\mu$m by Hony et al. (\cite{hony}). {\it Right:} 
The map simulated at 11.9 $\mu$m using the DUSTY code, assuming the parameters reported in table \ref{pardusty}.
The map has been convolved with a circular beam having HPBW of 0$^{\prime \prime}$.83.
In both the maps the contours indicate the 5-95\% intensity levels in steps of 10\%.}
\label{ir07}
\end{figure*}

For sources with only an upper limit on IRAS 100 $\mu$m flux, the same analysis as above was performed using the IRAS 60 $\mu$m flux to constrain minimum value for $p$. In fig.~\ref{bbody} 
the resulting curves are overplotted to the observed SEDs.

In table~\ref{masstab} the best fit parameters are listed together with the derived dust opacities and masses
calculated from the eq.~\ref{eq1} for $\lambda$ = 1.2 mm.

\subsection{SED modeling with DUSTY code}
\begin{figure*}
\resizebox{17cm}{!}{\includegraphics{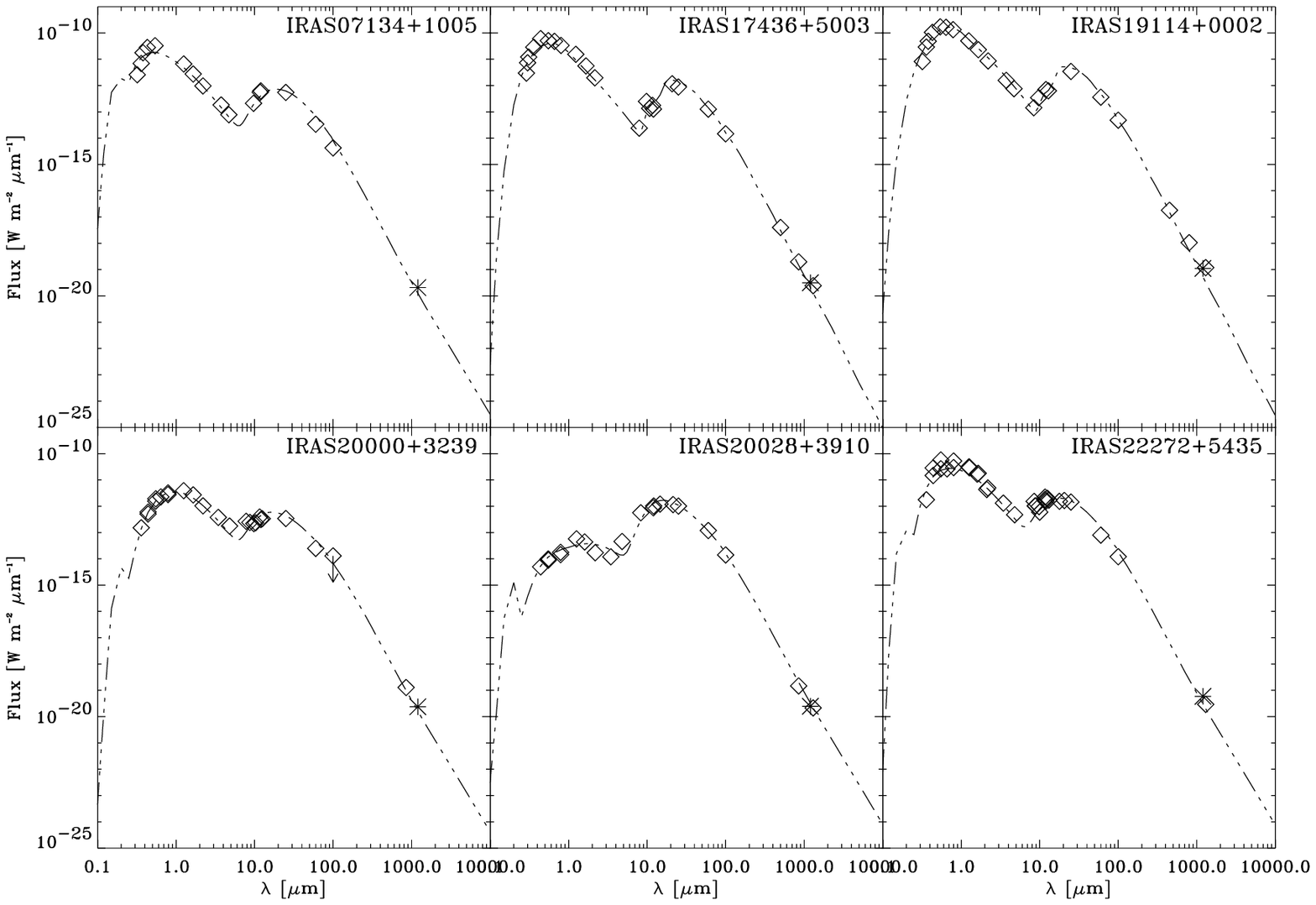}}
\caption{Spectral energy distributions of the emission towards six stars in the sample.
The asterisks indicate measures from this work.
The dashed lines show the SEDs computed using the DUSTY code. }
\label{fit}
\end{figure*}

For six
of the detected post-AGB stars, the observations scan a wide spectral range 
between optical bands and radio regions.
All the available informations on the sources, 
including the results from SED fits with envelope models, have been collected from the literature,
in order to verify the compatibility of the previously determined parameters with our 1.2 mm fluxes.

%
%
The observed SEDs are plotted in fig. \ref{fit}, along with best fit SEDs from
the radiative transfer code DUSTY (Ivezi\`{c} et al. \cite{ivez}). The code allows to calculate the
radiative transfer through a spherically symmetric dust shell and determines the
spectral energy distribution on the basis of specified properties of the radiation source, 
dust composition and dust shell properties. 
In particular, DUSTY allows to use six different types of dust grains 
which are assumed to be distributed in a spherical shell. For all the sources we assumed  
Planckian SED for the central star and the modified MRN (Mathis, J.S., Rumpl, W., Nordsieck, K.H. \cite{mathis}) power law
for the grain size distribution $n(a) \propto a^{-q}$, where $a$ is the grain size and 
{\it q} is fixed to 3.5, as it is commonly used, if not otherwise specified in the text.
For more details about the other options
we reference to the description in the user manual. When available, we calculated the SED
by assuming the stellar and dust parameters taken from previous references in literature,
otherwise the best-fit model parameters are sought iteratively by fitting the shape
of the observed SED.
 With the aim of add more observative constrains, we compared the infrared maps of the sources
available in literature with the simulated maps obtained from the code at the same observative frequencies.
{\bf All the simulated maps have been convolved with the beam of the instrument used to obtain the
map used for comparison}.
An example of such a comparison is shown
in fig. \ref{ir07} in the case of IRAS~07134+1005.
 In particular we refer to Hony et al. (\cite{hony})
for IRAS~07134+1005, Gledhill et al. (\cite{gled03}) for IRAS~17436+5003, Jura \& Werner (\cite{jura99}) for
IRAS~19114+0002, Ueta et al. (\cite{ueta01}) for IRAS~22272+5435 and to 2MASS Atlas Image provided from
Infrared Science Archive (bungo.ipac.caltech.edu/applications/2MASS/IM/ interactive.html) for
IRAS~20000+3239 IRAS~20028+3910.
We thus rejected
all those solutions that involve shell sizes which don't match the observative constrains according
to the rms noise level of the measures.

The remaining four sources of our sample, detected in the survey, were not fitted because of the lack of
either data and knownledge of stellar parameters, involving too many degrees of freedom.

A summary of the adopted input parameters used to produce the model fits in fig. \ref{fit} is
given in table \ref{pardusty}. In addition, we report the
derived stellar and dust shell parameters
beside the physical quantities used in scaling the dusty outputs 
in accordance with the dusty manual (Ivezi\`{c} et al. \cite{ivez}).
In all calculations we assumed a density for the grain material of 3 g/cm$^3$ and
a gas/dust ratio of 220.\par
The fit's parameter that is mainly constrained from the millimetric and submillimetric
measures is the shell outer radius, as these observations probe the cool, outer parts of the
CSE. However, the value derived from the SED fit, and reported in table \ref{pardusty}, has to
be meant as an estimate at the distance from the star where the wind density is close 
to the ambient density. Thus, the modeled envelope could result to be more extended than 
the observed one if its outer parts comes to have surface brightness lower or equal to
the rms noise level. In particular, it is not surprising that the radii derived for
the CSE outer edge result to be much greater than those derived from the near or mid
infrared observations, since at those frequencies most of the flux comes from the hotter
inner regions of the envelope.

{\bf On the basis of the mass loss rate and shell size derived from the SED fit, we calculated the
total envelop mass, assuming a constant wind expansion velocity which has been taken from
references in literature. 
The dust envelope masses derived for a gas/dust ratio of 220 are 
also reported in table~\ref{pardusty}.
Such values are consistent with those reported in table~\ref{masstab}  within a factor of 2.
We can thus asses that there is
a good agreement between the two dust mass estimates, expecially given the rough approximations made 
in the assumption of constant mass loss rate and in the choice of the gas/dust ratio, which could 
differ significatively from the assumed value as a function of the C/O ratio.}

The fit of the model to the
observed flux distribution for each object is discussed in the next section.

\section{Results and discussion}
{\it IRAS~07134+1005}. This object is known to have strong features near 21 and 30 $\mu$m, whose carrier species have not been firmly identified (Kwok et al. \cite{kwok89}). It is an F5 supergiant having C/O $\approx$ 1
(Van Winckler \& Reyniers \cite{vanwin00}). Despite of the very simplified dust composition used in our model calculations, 
the parameters derived from our fit are in good accordance  with the ones calculated from Meixner et al. \cite{meix04} 
and Hony et al. (\cite{hony}), using dif\-fe\-rent dust radiative transfer programs, with  the exception of the 
mass loss rate that is larger for a factor of 3
from the one  calculated from Hony et al. (\cite{hony}). On the other hand, our derived value for $\dot{M}$ agrees with the values obtained from the SED model performed by Hrivnak et al. (\cite{hriv00}) and from Jura et al. (\cite{jura00}) on the basis of their mid-IR  images of the source.\\
{\it IRAS~17436+5003}. It is a  F3 Ib supergiant with an oxygen-rich chemistry (Justtanont et al. \cite{justt}). Its SED has been modeled by several authors adopting different choices mainly in the selection of
grain parameters. While Hoogzaad et al. (\cite{hooz}) and Meixner et al. (\cite{meix02}) adopt the usual grain size distribution
$n \propto a^{-3.5}$ with a minimum size of 0.18 $\mu$m and 0.2 $\mu$m respectively, Gledhill \& Yates (\cite{gled03}) opt
for very small grains, using a steep size distribution $n \propto a^{-6}$ and minimum grain size 0.01 $\mu$m,
that, as pointed out from the authors, better account for both the SED shape within the whole frequency
range and the observed high degrees of near-IR linear polarization. In our calculation we followed the latter
way. As a consequence, the dust mass reported in tab.~\ref{masstab} could be underestimate 
because the assumed dust opacity is derived from an approximated grain model with an average grain size of $0.1$ $\mu$m
( Hildebrand, \cite{hildeb}).
This value 
is a factor of 5 greater than the grain size obtained from the $n \propto a^{-6}$ distribution in the adopted range for $a$. Assuming 
for such small grain radius $\chi_{850\mu\mathrm{m}} = 0.54$ cm$^2$ g$^{-1}$ (Gledhill \& Yates \cite{gled03}), the derived
dust mass increases of one magnitude order.

\begin{table*}
\caption[]{Input and derived stellar and dust parameters resulting from the fit to the SED.}
\label{pardusty}
\begin{center}
\begin{tabular}{c c c  c  c  c  c  c }
  \hline
   \hline
Property &IRAS	   & 07134+1005 & 17436+5003 & 19114+0002 & 20000+3239 & 20028+3910 & 22272+5435 \\
	   \hline	   
T$_{\ast}$[K] && 7250  & 7500  & 5660 & 5000  & 7000  & 5300  \\
Chem.      &&  C    &    O  &  O   &  C   &  C?   & C     \\
T$_d$[K]   &&  150  &   120  & 110 & 170  &  180  &  160  \\
L [L$_{\odot}$]  && 6000 & 3440 & 300000 & 630 & 6600 & 6700 \\
 A$_V$       &&  0.5  &  1.2 & 2   &  2.5  &  1.3 & 2.5 \\ 
 r$_{i}$ [cm] && $4.76\times 10^{16}$ & $1.91\times 10^{16}$ &  $1.73\times 10^{17}$ & $8.18\times 10^{15}$ & $1.88\times 10^{16}$ & $2.46\times 10^{16}$\\
r$_{o}$ [cm] && $3.48\times 10^{17}$ & $1.53\times 10^{17}$ &  $6.88\times 10^{17}$ & $2.05\times 10^{17}$ & $1.88\times 10^{17}$ & $3.11\times 10^{17}$\\
 $\dot{M}$ [$\frac{M\odot}{yr}$] &&   $3.48\times 10^{-5}$ & $3.48\times 10^{-5}$ & $1.09\times 10^{-3}$ & $6.92\times 10^{-6}$ & $8.93\times 10^{-5}$ &  $3.12\times 10^{-5}$   \\ 
v$_{exp}$ [km/s] && 10 .0 & 15.5 & 35.0 & 12.0 & 16.0 &  10.0 \\
M$_{d}$ [M$_{\odot}$] && $1.66\times 10^{-3}$ & $4.35\times 10^{-4}$ & $2.31\times 10^{-2}$ & $1.63\times 10^{-4}$ & $1.35\times 10^{-3}$ & $1.28\times 10^{-3}$\\
 Ref. &  &  {\scriptsize 1, 2, 3, 4}   &{\scriptsize 5, 6, 7, 18, 19, 20} & {\scriptsize 3, 8, 9, 10, 18, 19, 21}  & {\scriptsize  11, 12, 13, 19, 22} & {\scriptsize 14, 15, 18, 19, 20} & {\scriptsize 13, 16, 17, 18} \\
 \hline
\end{tabular}
\\
\end{center}
\footnotesize{References - (1) Meixner et al. \cite{meix04}; (2) Hony et al. \cite{hony}; (3) Hrivnak et al. \cite{hriv89}; (4)Van Genderen et al. \cite{vangen}; (5) Ghedill et al. \cite{gled03}; (6) Ueta et al. \cite{ueta00}; (7) Skinner et al. \cite{skinner}; (8) Van der Veen \cite{vanderveen}; (9) Hawkins et al. \cite{hawkins}; (10) Th\'evenin et al. \cite{thevenin}; (11) Kwok et al., \cite{kwok95}; (12) Hrivnak et al. \cite{hriv95}; (13) Volk et al. \cite{volk}; (14) Su et al. \cite{su}; (15) Bujarrabal et al. \cite{bujar01};
(16) Hrivnak et al. \cite{hriv91}; (17) Ueta et al. \cite{ueta01}; (18) Walmsley et al. \cite{walmsley}; (19) Gledhill et al. \cite{gled02}; {\bf (20) Likkel et al. \cite{lik91}; (21) Bujarrabal et al. \cite{bujar92}; (22) Hrivnak et al. \cite{hriv00}}.}
\end{table*}

The inner shell radius derived from the SED fit agrees with the value of 1400 AU ($2.1\times 10^{16}$ cm at 1.2 kpc) obtained
by Hoogzaad et al. (\cite{hooz}), but is about 30 -- 40\% larger than the same parameter derived from Gledhill \& Yates (\cite{gled03}) and Meixner et al. (\cite{meix02}), who nevertheless used an asymmetric dust model.
The outer radius is larger than the values obtained from  Hoogzaad et al. (\cite{hooz}) and Meixner et al. (\cite{meix02}), but close to that derived by Gledhill \& Yates (\cite{gled03}) and to the extension of
 6$\arcsec$.5 derived from the CO observation (Bujarrabal et al., \cite{bujar92}).\\
{\it IRAS~19114+0002}. It is an oxygen-rich object of spectral type G5 Ia (Hrivnak et al. \cite{hriv89}),
but the real evo\-lu\-tio\-na\-ry status of this object is still controversial. It has been
classified either as post-AGB star having a luminosity of about $10^4$ L$_{\odot}$ and lying at 1 kpc (Hrivnak et al. \cite{hriv89}) or as a massive red supergiant with luminosity near $3 \times 10^5$ L$_{\odot}$ and lying to a distance of 6 kpc (Hawkins et al. \cite{hawkins}).

The expansion velocity of the circumstellar envelope $v_\mathrm{e}\approx 35$ km s$^{-1}$ determined from the  profile of the circumstellar CO emission (Zuckerman \& Dyck \cite{zucker}; Bujarrabal et al. \cite{bujar92}) is significantly higher than the typical value of 15 km s$^{-1}$ for an AGB star and favors the supergiant hypothesis, that is though still doubt (Josselin \& L\`ebre \cite{bujar92}).

As for IRAS~17436+5003, Gledhill \& Takami (\cite{gled01}) pointed out the need to adopt a steep power law for the grain size distribution, in order to agree with the high degrees of linear polarization observed in the near-IR. We have thus adopted the grain size distribution $n \propto a^{-6}$ {\bf with $a_{\mathrm{min}}$= 0.005 $\mu$m and $a_{\mathrm{max}}$= 0.25 $\mu$m. The SED fit gives a $r_{\mathrm{in}} = 1.7\times 10^{17}$ cm, which is in good accordance with the inner shell radius of
about $1.5\times 10^{17}$ cm obtained from both mid-IR images (Jura \& Werner  \cite{jura99}) and near-IR imaging polarimetry observations (Gledhill et al. \cite{gled01}). Such value is very close to the inner shell size of $1.8\times 10^{17}$ cm derived, for the assumed distance of 6 kpc, from high resolution CO observations performed from Jura et al. (\cite{jura01})}. 

The dust mass obtained is in good accordance with the one derived from Gledhill et al. (\cite{gled02}) on the basis of submillimeter observations.
\\
{\it IRAS~20000+3239}. Low resolution K-band spectra performed by Davis et al. (\cite{davis}) showed a rather compact source
with angular size lower than 2$\arcsec$, as well Hrivnak et al. (\cite{hriv99}) measured 1$\arcsec$.6 in diameter in V band. 
Comparing with the SED parameters derived from Volk et al. (\cite{volk}), we obtain good accordance for both the mass loss rate and inner shell radius, as well as for the shell mass derived from Gledhill et al. (\cite{gled02}).\\
{\it IRAS~20028+3910}. This object is characterized from a bipolar morphology, with the central object highly
obscured in optical and near-IR (Su et al. \cite{su}; Ueta et al. \cite{ueta00}). The SED, constructed with the
data reported by Su et al. (\cite{su}) and references therein, thus shows a mid and far infrared peak much most bright than
the one in the near infrared. Neri et al. (\cite{neri}) fitted the CO 1-0 visibility data with an elliptical gaussian component with a
size of $3.5\arcsec \times 11.1\arcsec$. The dust mass obtained from our 1.2 mm measures agrees very well with the value calculated from previous
submillimeter data (Gledhill et al. \cite{gled02}).\\
{\it IRAS~22272+5435}. It is an extremely carbon rich  object and, as IRAS~07134+1005, shows peculiar infrared spectral 
features. From the analysis of CO 1-0 visibility data, Neri et al. (\cite{neri}) measured an extended envelope size (FWHM) of 21$\arcsec$. 
From subarcsecond mid infrared ima\-ging study (Ueta et al. \cite{ueta01}) the dust shell was found to have a toroidal structure with a 0.5$\arcsec$ inner radius, which corresponds to 1.1$\times$10$^{16}$ cm at 1.6 kpc. Our fit indicates an inner radius which is consistent
with that measured by Ueta et al. (\cite{ueta01}) to whithin a factor of 2. A major discrepancy is found in the estimates of mass loss rate.
On the basis of their radiative transfer calculations, in fact, Ueta et al. (\cite{ueta01}) derived a wind with 
$\dot{M} = 4.1 \times 10^{-6}$ M$_{\odot}$ yr$^{-1}$. Our value is closer with the mean mass
loss rate obtained from the CO study performed Bujarrabal et al. (\cite{bujar01}), 
which is $1.8 \times 10^{-5}$ M$_{\odot}$ yr$^{-1}$, 
when scaled for our assumed distance. The authors report a total nebular mass of 0.18 M$_{\odot}$, 
that gives a dust mass of
$8.0 \times 10^{-4}$ M$_{\odot}$ for an assumed gas to dust ratio of 220, and which is in good agreement with our estimate.

\section{Summary and outlook}
We have presented the results of 1.2 mm continuum observations for a sample of 24 sources classified as post-AGB.
Continuum emission was detected toward 11 objects, while uncertain detection is reported for two other sources.\par
The circumstellar dust masses were derived from our 1.2~mm measures, assuming that the emission is due to optical thin 
dust. For the sources of which the distance is known, the circustellar dust masses range between about $6 \times 10^{-4}$~M$_\odot$ and $2.4 \times 10^{-3}$~M$_\odot$, with exception of IRAS~19114+0002, whose post-AGB nature is, however, still in question. For the other objects we derived lower dust masses, indicating a more young circumstellar envelope, but the errors in the choice of the common assumed distance could have effected the derived values.\par
For six of the detected sources, we compared the observed SEDs, constructed with additional data from li\-te\-ra\-tu\-re, and the model spectra obtained using a sym\-me\-tric radiative transfer code. This allowed us to estimate 
some physical parameters of stars and envelopes which have been compared with previously results reported in literature.

The high detection rate ($\approx 46\%$) seems to support the goodness of our selection rules and an extension of the millimetric survey to the remain targets in our sample is necessary. Stars  in our full  sample  belong to different evolutionary phases in the transition from AGB to PN,  as the stars with optical counterpart should be more evolved. Once the full sample will be observed, the comparison between the derived physical properties of different envelopes  will provide fundamental informations on this evo\-lu\-tio\-na\-ry phase  not fully understood yet.

We still note that the CSEs surrounding such kind of objects
appear as good targets for the first light projects for future millimeter arrays such as the Atacama Large Millimeter Array (ALMA).
From our fits we derive typical dimensions of CSEs ranging from $1.5 \times 10^{17}$ to  $1.2\times 10^{18}$ cm, which, combined with the distances as reported in table 3, correspond 
to angular sizes from few arcsecs up to $\approx 13\arcsec$. This implies that such CSEs can be, in principle, mapped, in great details  with the foreseen ALMA angular resolutions  by using a combination
of both compact (to fully recover all the flux) and extended con\-fi\-gu\-ra\-tions (Wilson et al. 2005). 
The foreseen capabilities of ALMA will allow, at least for the more compact sources of our sample,
and in general for post-AGB sources, to directly map the dusty envelopes at several millimetric and
submillimetric frequencies. This could provide a better constrains to the modeling of CSE. Furthermore, 
a detailed map of CSE could evidence any kind of structured morphologies that can be related to
different mass loss episodes, suffered by the star during the AGB evolutionary phase.\par
\begin{figure}
\resizebox{10cm}{!}{\includegraphics{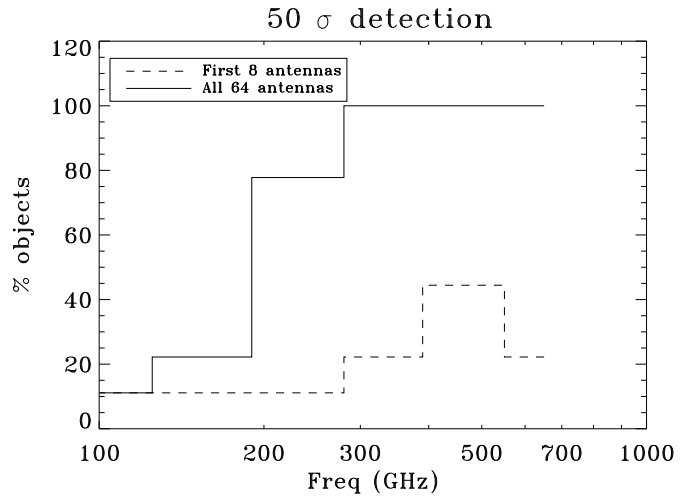}}
\caption{Percentage of objects, detected at 1.2 mm, which are observable with dynamical range grater than 50, assuming calculate ALMA sensitivity with both 8 and 64 antennae.}
\label{alma}
\end{figure}

To evaluate the possibility to actually resolve and map the CSEs of our sample, we need to compare, at each frequency channel, the expected flux densities with the foreseen ALMA sensitivity. 
From the fitted SEDs we have then extrapolated  the flux densities in the ALMA first light channels, namely 0.5, 0.6 0.9 1.3, 2 and 3 mm. Expected ALMA sensitivities have been calculated by using the ALMA Sensitivity Calculator 
(www.eso.org/projects/alma/science/bin/sensitivity.html) in the case of first-light, assuming that only 8 antennae will be available, and in the case of the full (64 antennae) array.
For both configurations we derived a detection rate close to $100$\% over the 3$\sigma$ 
at almost all frequencies.
In fig. \ref{alma} we show the expected percentage of the studied objects that ALMA will allow to observe with dynamical range greater than 50. 
It is evident how ALMA will allow us not only to better sample the millimetric range of the source SEDs, that up to now is very poor, but also to obtain, in most cases, multifrequencies high resolution maps of the circumstellar matter surrounding the stars of our sample, extended to the sources belonging to the south hemisphere.

\begin{acknowledgement}
 We  would like to thank the anonymous referee for his constructive criticism which
enabled us to improve this paper.
\end{acknowledgement}

\end{document}